\documentclass[%
 reprint,
 amsmath,amssymb,
 aps,
pra,
floatfix,
]{revtex4-2}
\usepackage{xcolor}
\usepackage{graphicx}
\usepackage{dcolumn}
\usepackage{bm}

\usepackage{hyperref}
\usepackage{soul}

\begin{document}

\preprint{APS/123-QED}

\title{Confining potential in holographic bottom-up QCD from WKB}
\author{Miguel Angel Martin Contreras}
\email{miguelangel.martin@usc.edu.cn}
\affiliation{
 School of Nuclear Science  and Technology,\\
 University of South China,\\
 Hengyang, China\\
 No. 28, West Changsheng Road, Hengyang City, Hunan Province, China.
}
\affiliation{Key Laboratory of Advanced Nuclear Energy Design and Safety,\\ Ministry of Education, Hengyang, 421001, China}

\author{Mitsutoshi Fujita}
\email{fujitamitsutoshi@usc.edu.cn}
\affiliation{
 School of Nuclear Science  and Technology,\\
 University of South China,\\
 Hengyang, China,\\
 No. 28, West Changsheng Road, Hengyang City, Hunan Province, China.
}
\affiliation{Key Laboratory of Advanced Nuclear Energy Design and Safety,\\ Ministry of Education, Hengyang, 421001, China}

\author{Alfredo Vega}%
 \email{alfredo.vega@uv.cl}
\affiliation{%
 Instituto de F\'isica y Astronom\'ia, \\
 Universidad de Valpara\'iso,\\
 A. Gran Breta\~na 1111, Valpara\'iso, Chile
}

\begin{abstract}
We solve the inverse quantum mechanical problem in holographic QCD by deriving a confining bulk potential directly from a given hadronic spectrum. Utilizing the semiclassical Rydberg-Klein-Rees (RKR) method, we reconstruct the potential corresponding to the vector meson spectrum of the top-down D3/D7 brane model. The derived bottom-up potential exhibits a sharp infrared cutoff, structurally mimicking the hardwall model. We characterize this new holographic model by computing the thermal deconfinement transition temperature and the configurational entropy of the $\rho$ meson trajectory. Our methodology offers a novel approach to reverse-engineer gravitational duals from spectroscopic data, thereby bridging the top-down and bottom-up approaches to confinement.
\end{abstract}
\maketitle
\section{\label{sec:intro} Introduction}
AdS/CFT correspondence \cite{Maldacena:1997re} since its original formulation has become an excellent tool to address non-perturbative phenomena beyond gravitational physics. Over the past twenty-five years, a plethora of effective models have been developed, which can be broadly classified into two main groups based on their construction: top-down and bottom-up. 

Top-down models, roughly speaking, are constructed \emph{systematically} from a well-defined string theory, thereby yielding a holographically consistent gauge theory. In other words, we are dealing with an entire gauge/gravity duality framework, assuming complete knowledge of both sides, thereby allowing exact predictions based on known theories. Therefore, the top-down approach is mainly used to test and refine the AdS/CFT correspondence.  

Conversely, the bottom-up approach studies and understands AdS/CFT by starting with specific, well-defined components and building a more comprehensive understanding. This customization offers more flexibility than the top-down approach. However, the payoff of such an approach is that bottom-up approaches are effective models, i.e., they are restricted by their phenomenological framework. A good example is the plethora of research on AdS/QCD. 

In the particular case of holographic QCD, both perspectives have been used to address hadronic spectroscopy. However, the mechanism that induces confinement, \emph{ergo} the emergence of bound states, varies across paradigms. In the top-down case, deforming the geometry naturally induces an energy scale that sets the hadron masses \cite{Erdmenger:2007cm}. In the bottom-up case, confinement is induced by breaking the bulk conformal symmetry, while preserving the AdS conformally flat limit exhibited by the bulk modes dual to hadrons at the boundary. However, this procedure can be performed by deforming the AdS space \cite{Polchinski:2002jw, Boschi-Filho:2004rrk, Erlich:2005qh, Forkel:2007cm} or by inducing a dilaton field \cite{Karch:2006pv}. Both scenarios allow the emergence of bulk-bound states dual to hadrons.  

In the case of the bottom-up proposal, the engineering is summarized as follows: given a dilaton field or mea metric deformation, it is possible to write down a confining potential that yields a bound spectrum when solving a Schrödinger-like equation associated with it. We call this approach \emph{direct Schrödinger problem.}

On the other hand, it is possible to formulate the \emph{Schrödinger inverse problem}, where the starting point is the bound spectrum used to compute the confining potential. This inverse problem has been studied in the framework of molecular quantum physics using the WKB method, in which the confining potential can be derived from the energy spectrum. This procedure is known as the Rydberg \cite{Rydberg1, Rydberg2} -- Klein \cite{Klein} -- Rees \cite{Rees} (RKR) method. Notice that to use RKR, we need to consider \emph{smooth and monotonically increasing (near to the turning points) potentials} and the spectrum should not be \emph{noisy} or degenerate \cite{Quigg:1979vr}.

To illustrate and validate this methodology in a controlled setting, we require a well-defined, theoretically consistent spectrum. We therefore consider the vector meson Regge trajectory derived from the top-down D3/D7 system \cite{Kruczenski:2003be}, which provides a clean input spectrum free from experimental uncertainty. Thus, using the RKR formulas, we obtain a bottom-up associated potential with its corresponding dilaton. The reconstructed bottom-up potential resembles the hardwall model \cite{Polchinski:2001tt}. 

This work is organized as follows. Section \ref{hadrons} briefly introduces the bottom-up description of hadrons. Section \ref{WKB} introduces the concepts underlying the RKR formulas for reconstructing potentials from Regge boundary trajectories. As an illustrative example, we reconstruct the static quadratic dilaton, i.e., the softwall model \cite{Karch:2006pv}. 
Section \ref{WKB-Bottom-up} presents the fundamental concepts of the D3/D7 system and explains how the meson masses are generated within this framework. Subsequently, we utilize the D3/D7 Regge trajectory as input data to compute a bottom-up confining potential.
Section \ref{thermal} discusses the confinement/deconfinement phase transition, following the well-known analysis by Herzog \cite{Herzog:2006ra}. Section \ref{DCE} addresses the configurational entropy for vector mesons computed within the reconstructed bottom-up framework. Finally, we present our conclusions in Section \ref{concl}.

\section{Mesons in bottom-up models}\label{hadrons}
In the bottom-up approach, mesons are defined using the normalizable bulk modes that emerge when a confinement scale breaks conformal invariance in the bulk. We use a dilaton field $\Phi(z)$ that can be static or dynamically generated. Along with the dilaton, we must establish a geometric framework. It is customary to work on the \emph{Poincaré Patch} defined as

\begin{equation}\label{AdS-metric}
    dS^2=\frac{R^2}{z^2}\left(dz^2+\eta_{\mu\nu}\,dx^\mu\,dx^\nu\right)=g_{MN}\,dx^M\,dx^N,
    \end{equation}

\noindent where $R$ is the AdS radius, $z$ is the so-called \emph{holographic coordinate} and $\eta_{\mu\nu}$ is the Minkowski metric that describes the conformal boundary at $z\to0$. We use Capital Latin indices for 5-dimensional tensor objects and Greek ones for their 4-dimensional Minkowskian parts.

The bulk action in this kind of model is given as usual, as 

\begin{equation}\label{action-mes}
    I_\text{Mesons}=\int{d^5x\,\sqrt{-g}\,e^{-\Phi(z)}\,\mathcal{L}_{\text{Mesons}}},
\end{equation}

The Lagrangian density for the 1-form bulk fields (dual to vector mesons) can be summarized as

\begin{equation}\label{lagrangian}
    \mathcal{L}_\text{Mesons}=-\frac{1}{2\,g_5^2}\,\nabla_M\,\phi^{N}\,\nabla^M\,\phi_{N}+\frac{1}{2}\,M_5^2\,\phi^{N}\,\phi_{N},
\end{equation}

\noindent with $g_5$ a scale that fixes units in the action and also helps to define the decay constants \cite{Erlich:2005qh}. The bulk mass $M_5$ plays an important role in the bottom-up approach. This quantity defines the \emph{hadronic identity}  associated with the bulk modes. From the field/operator duality, the dimension $\Delta$ of operators creating mesons, composed of quark ($Q$)-antiquark ($\bar{Q}$ pairs,  at the boundary is dual to the scaling of the bulk fields in the limit $z\to 0$.

The most general meson operator at the boundary can be defined as $\mathcal{O}_{3+l}=\bar{Q}\, D_{\{l_1}\dots D_{l_k\}}Q$, written in terms of the symmetrized products of the gauge covariant derivative $D_\mu$. The operator $\mathcal{O}_{3+l}$ has total spacetime orbital momentum, $l=\sum_{i=1}^{k}\,l_i$ and conformal dimension $\Delta=3+l$ \cite{deTeramond:2005su}. Notice that each quark carries an energy dimension of $3/2$, and every covariant derivative is dimension one.

In this scenario, the bulk mass is defined as  

 \begin{equation}
     M_5^2\,R^2 =\frac{1}{4}\left(3+2\,l-\beta\right)\left(1+2l+\beta\right)
 \end{equation}
 
\noindent where $\beta=-1$ characterizes the vector meson spin. In the holographic QCD approach, we consider that the bulk $p$-form index corresponds to the hadron spin $J$, with $J=\left|l+S\right|$, in agreement with the field/operator duality, where $S$ is the $Q\bar{Q}$ spin contribution to the total meson spin.

From variations in the  bulk action \eqref{action-mes}, we can obtain the equations of motion for the bulk vector field $\phi_M$ as

\begin{multline*}
    \frac{1}{\sqrt{-g}}\,\nabla_R\left[\sqrt{-g}\,e^{-\Phi}\,g^{RN}\,g^{M M'}\,\nabla_N\,\phi_{M'}\right]-\\
    M_{5}^2\,e^{-\Phi}
    \,g^{M M'}\,   \phi_{M'}=0.
\end{multline*}

Notice that these equations are valid for the \emph{weak gravity limit} only, which considers that the affine Levi-Civita connections are zero (but not their derivatives) \cite{Gutsche:2011vb}. At this step, we Fourier transform and decompose the bulk field as $\phi_M=\tilde{\phi}_M(q)\,\psi(q,z)$ where $\tilde{\phi}_M(x)$ is the boundary source and $\psi(q,z)$ is the bulk mode. For normalizable solutions, we impose the boundary condition $\psi\sim z^{\Delta-1}$ \cite{Aharony:1999ti}. Thus, the bulk equations of motion reduces to \emph{Sturm-Liouville} equation for $\psi(q,z)$

\begin{multline}\label{sturm-liouville}
    \partial_z\left[e^{-B(z)}\,\partial_z\,\psi(z)\right]+\left(-q^2\right)\,e^{-B(z)}\,\psi(z,q)\\
    -\frac{M_5^2\,R^2}{z^2}\,e^{-B(z)}\psi(z,q)=0
\end{multline}

\noindent where we have defined $B(z)=\Phi(z)+\beta\,\log\left(\frac{R}{z}\right)$ and $-q^2=M_n^2$ is the on-shell mass condition. We also have imposed the gauge fixing $\phi_z=0$, which is consistent with the transverse condition $\partial_\mu\,\phi^\mu =0 $. 

The meson spectrum $M_n^2$ emerges from the equation above. However, it is customary to transform the Sturm-Liouville equation into a Schrödinger-like equation that is characterized by the emergence of a \emph{holographic potential} whose eigenspectrum is the Regge trajectory for mesons characterized by the dilaton field $\Phi(z)$.

To do so, we define the \emph{Bogoliubov transformation} 
 $\psi(z)=e^{\frac{1}{2}B (z)}\,u(z)$ such that the Sturm-Liouville equation reduces to the expected Schrödinger-like equation:

\begin{equation}\label{eq-sch-like}
    -u''(z)+V(z)\,u(z)=M_n^2\,u(z),
\end{equation}

with the holographic potential defined in terms of the dilaton and the AdS warp factor as 

\begin{multline}\label{holographic-pot}
  V(z)=\frac{M^{2}_{5}\,R^{2}}{z^{2}}-\frac{\beta(2-\beta)}{4z^{2}}\\ -\frac{\beta}{2\,z}\Phi^{\prime}(z)
+\frac{1}{4}\Phi^{\prime}(z)^{2}-\frac{1}{2}\Phi^{\prime\prime}(z).
\end{multline}

By solving this potential, we obtain the mass spectrum. This potential will serve as the starting point for the next section.

\section{WKB reconstruction}\label{WKB}
One possible approach to the inverse Schrödinger problem is to use the WKB method. Suppose we have a hadronic Regge trajectory, either from experimental fits or proposed by non-holographic models, that can be written as a function of the excitation number, i.e., $M_n^2(n)$. 

The key question in this inverse problem is what phenomenology is dictated by the large-$z$ behavior of the potential. This limit will determine the corresponding large-$z$ dilaton profile. In the context of Regge trajectories, this information suffices to describe them. The large-$z$ behavior of the holographic potential governs the linearity and slope of the Regge trajectory. Conversely, the low $z$ limit yields the intercept. 
 
We learned from the original softwall \cite{Karch:2006pv} that the holographic potential large-$z$ behavior naturally controls confinement. In the case of top-down approaches, the equivalent of large $z$ behavior \emph{conveniently} deforms the bulk space and bulk fields to induce a mass of excited states dual to mesons, as in the D3/D7 approach \cite{Kruczenski:2003be}.  

The main idea is to extract the static dilaton from the radial Regge trajectory. To do so, we will consider a general trajectory defined as

\begin{equation}\label{test-trajectory}
    M_n^2(n)=f(a_i,n),
\end{equation}

\noindent where $a_i$ defines a set of parameters, i.e., energy scales, that fix the units in the trajectory and $n$ stands for the radial quantization number. For example, for heavy mesons, it is expected that trajectories do not exhibit linearity \cite{Chen:2018nnr}. For certain heavy-light systems, large distances in the Cornell-like potential imply the emergence of a Coulomb-like spectrum \cite{Chen:2018hnx}. 

As explained in Appendices \ref{app-1} and \ref{app-2}, the large
$z$ behavior of the potential near the turning points can be inferred using the WKB approximation through one of the formulas of the RKR method. Since the conformal boundary is at $z\to0$, for holographic potentials the turning points are $z\to0$ and $z(V^*)$ with the property that $z(V^*)\to\infty$, thus we have 

\begin{equation}\label{turning}
z(V^*)=2\int_0^{V^*}{\frac{d\,M^2}{\frac{d\,M^2}{d\,n}\left(V^*-M^2\right)^{1/2}}}.     
\end{equation}

Inverting this expression, we find the high-$z$ potential responsible for confinement in bottom-up models. The near-conformal boundary behavior of the potential and dilaton controls the decay constants, as was shown in refs. \cite{Braga:2017bml, MartinContreras:2019kah}.  We use $V^*$ to denote the potential calculated using WKB from the full bottom-up holographic potential $V(z)$, which can be written as

\begin{equation}
    V(z)=\frac{\beta(\beta-2)+4\,M_{5}^2\,R^2}{4\,z^2}+V^*(z),
\end{equation}

\noindent where the hadronic identity is encoded in the bulk mass $M_5$. This procedure has a fundamental limitation inherent to semiclassical inversion methods. The reconstructed potential is guaranteed to reproduce the input spectrum accurately only in the semiclassical limit of large radial excitation number $n$. Consequently, the equivalent bottom-up potential does not exactly match the input spectrum for low-lying states. The ground state, for instance, could deviate slightly ($2.7\%$ for the case of the $\rho$ meson, as described in Table \ref{tab:one}). While high precision is expected for excited states, reproducing the low-$n$ spectrum exactly would require additional fine-tuning beyond the RKR procedure.

Once we have the equivalent bottom-up potential from the spectrum, we can calculate the associated dilaton field $\Phi(z)$. To do so, we use reverse engineering on the potential $V^*(z)$ as follows:

\begin{equation}\label{dilaton-eng}
    V^*(z)=\frac{1}{4}\Phi'(z)^2-\frac{1}{2}\Phi''(z)-\frac{\beta}{2\,z}\Phi'(z).
\end{equation}

At low $z$, the holographic potential is not sensitive to the dilaton since, in this region, $\Phi(z\to0)$ is expected to converge softly to a constant value. 

\subsection{softwall model}
The softwall model is characterized by linear Regge trajectories, i.e., $M^2(n) = a(n+b)$. In this case (\ref{turning}) produces:

\begin{equation}
z(V^*)=2\int_0^{V^*}{\frac{d\,M^2}{a \left(V^*-M^2\right)^{1/2}}} = \frac{4}{a} \sqrt{V^{*}}.  \end{equation}

Therefore, the large $z$ differential equation has the following structure: 

\begin{equation}
    \frac{1}{16}\,a^2\,z^2=\frac{1}{4}\Phi'(z)^2-\frac{1}{2}\Phi''(z)+\frac{1}{2\,z}\Phi'(z),
\end{equation}

\noindent where we have fixed $\beta=-1$ for the vector mesons. 

Solving the equation above, we obtain for the dilaton the following expression:

\begin{equation}
\Phi(z)=c_1-2\,\text{log}\left[\text{cosh}\left(\frac{a}{8}z^2-2\,c_2\right)\right].
\end{equation}

For the argument inside the logarithm function, we can write: 

\begin{equation}
 \text{cosh}\left(\frac{a}{8}z^2-2\,c_2\right)=\text{cosh}\,\frac{a}{8}z^2\,\text{cosh}\,2\,c_2-\text{sinh}\,\frac{a}{8}z^2\,\text{sinh}\,2\,c_2.
\end{equation}

Fixing $c_1=0$ and $c_2\to\infty$, implying that $\text{cosh} 2\,a\,c_2=\text{sinh}\,2\,a\,c_2=1$. Therefore, we obtain an expression for the dilaton field coming from boundary information as follows

\begin{equation}
\Phi(z)=-2\,\text{log}\left(e^{-\frac{a}{8}z^2}\right)=\frac{a}{4}z^2.
\end{equation}

If we compare with the standard softwall model, where $\Phi(z)=\kappa^2\,z^2$, we can infer the dilaton slope in terms of the Regge slope as $\kappa^2=a/4$. This fact immediately implies that we recover the well-known result coming from linear confinement for the vector Regge trajectories at $S$-wave:

\begin{equation}
    M_n^2=4\,\kappa^2(n+1).
\end{equation}

Thus, we have recovered the softwall model. The results for scalar mesons can be obtained similarly by fixing $\beta =- 3$.

\section{Bottom-up approach to D3/D7 model}\label{WKB-Bottom-up}
Let us apply these WKB ideas to the D3/D7 model using a bottom-up approach~\cite{Karch:2002sh, Kruczenski:2003be, Ammon:2015wua}. Two types of D-branes in this model are considered non-perturbative solitonic objects, characterized by their tension and energy being inversely proportional to the string coupling constant, denoted as $1/g_s$, unlike fundamental strings  

The D3/D7 brane intersection is defined as a stack of $N_c$ coincident D3-branes,
embedded into the world volume of $N_f$ D7-branes \cite{Kirsch:2005uy}. These D7-branes are embedded such that they share the four spacetime directions with the D3-branes and extend along four of the six transverse directions, specifically directions $4$, $5$, $6$, $7$, as shown in Table  \ref{tab:zero}.

\begin{table}[h]
    \centering
    \begin{tabular}{c|c|c|c|c|c|c|c|c|c|c|}
            & $0$ & $1$ & $2$ & $3$ & $4$ & $5$ & $6$ & $7$ & $8$ & $9$\\
    \hline
        D3: &  $\times$ & $\times$ & $\times$ & $\times$&  & & & & &\\
        D7: & $\times$ & $\times$ & $\times$ & $\times$ & $\times$ & $\times$ & $\times$ & $\times$ & & \\
    \end{tabular}
    \caption{The D3/D7-brane system in $9 + 1$-dimensional flat space.}
    \label{tab:zero}
\end{table}

In the appropriate decoupling limit and requiring $g_s\, N_c\gg1$, the stack of D3-branes can be replaced with the AdS$_5\times$S$_5$ geometry with the AdS throat radius $R^2=\sqrt{4\,\pi\ g_s \ N_c}\alpha'$. If, in addition, condition $N_c\gg N_f$ is imposed, it is possible to neglect the backreaction of D7-branes, which implies that in the gravity description, they appear as $N_f$ D7-brane probes. This process entails breaking the original $\mathcal{N}=4$ supersymmetry down to $\mathcal{N}=2$, allowing the inclusion of dynamical quark fields analogous to those in QCD. Recall that quark fields transform in the fundamental representation of the gauge group, while $\mathcal{N}=4$ theory only contains fields in the adjoint representation. When D3 and D7 overlap, the quarks become a massless state. Specifically, the D7-branes are separated from the D3-branes along a perpendicular direction. $L$ is the distance between the D3 and D7 branes that sets the mass scale of the light state (quark fields). Thus, the $\mathcal{N}=2$ theory acquires a spectrum of mesons. See Refs. \cite{Kruczenski:2003be,Erdmenger:2007cm} for further details.

Geometrically speaking, the mesons in this model arise as fluctuations of the associated DBI action. These D7-brane plane wave fluctuations form a set of massive gauge supermultiplets of the $\mathcal{N}=2$ theory. We can identify the modes with mesons as those with conformal dimension $\Delta=l+3$. For these modes, the mass spectrum has the following structure: 

\begin{equation}
    M_{n,l}^2=\frac{4\,L^2}{R^4}\left(n+l+1\right)\left(n+l+2\right)
\end{equation}

The mass spectrum is dominated entirely by the large-$z$ terms in the potential. Thus, from WKB, we can infer the bottom-up counterpart of this D3/D7 spectrum. Applying the WKB integral, we have:

\begin{eqnarray}\notag
    z(V^*)&=&2\int_0^{V^*}{\frac{d\,M_{n,l}^2}{\frac{\partial\,M_{n,l}^2}{\partial\,n}\left(V^*-M_{n,l}^2\right)^{1/2}}}\\ \notag&=&2\int_0^{V^*}{\frac{d\,M_{n,l}^2}{a\,\sqrt{1+\frac{4\,M_{n,l}^2}{a}}\sqrt{V^*-M_{n,l}^2}}}\\
    &=&\frac{2}{\sqrt{a}}\,\tan^{-1}\left(2\,\sqrt{\frac{V^*}{a}}\right),
\end{eqnarray}

\noindent where we have defined the energy scale $a=4\,L^2/R^4$ that sets the meson masses. 

Inverting the equation above, we obtain the large $z$ part of the bottom-up potential as

\begin{equation}
    V^*(z)=\frac{a}{2}\tan^2 \left(\frac{\sqrt{a}\,z}{2}\right).
\end{equation}

Therefore, the bottom-up potential associated with the D3/D7 mass spectrum for mesons acquires the following structure:

\begin{equation}\label{D3_D7-pot}
    V_\text{D3/D7}(z)=\frac{\left(2\,l+3\right)\left(2\,l+1\right)}{4\,z^2}+\frac{a}{4}\tan^2 \left(\frac{\sqrt{a}\,z}{2}\right).
\end{equation}

Let us dissect this potential. First, we focus our attention on functional behavior. The potential has a periodic structure controlled by the scale $a=4\, L^2/R^4$. Thus, the potential will go to infinity when 

\begin{equation}
\label{cutoff}    z_\text{cutoff}=\frac{\pi}{\sqrt{a}}\,\left(2\gamma+1\right)=\frac{\pi\,R^2}{2\,L}\,\left(2\gamma+1\right)
,\,\,\,\gamma \in \left\{0,\mathbb{N}\right\}.
\end{equation}
Thus, $z_\text{cutoff}$ defines a \emph{wall} that prevents modes from going beyond these points. For simplicity, we will fix $\gamma=0$. This observation leads us to the second point. \emph{This confining potential has the same behavior as the one observed for the hardwall model} \cite{Boschi-Filho:2002wdj, Erlich:2005qh}. A phenomenological fact supports this claim. The hardwall model spectrum is defined by the zeroes of Bessel functions of the first kind as 

\begin{equation}\label{HW-mass}
    M_n^2=\left(\frac{\alpha_{1,n}}{z_\text{HW}}\right)^2,
\end{equation}

\noindent where the energy scale is fixed by the locus $z_\text{HW}$ of the D-brane used to cut the AdS slice. When we consider the asymptotics of $\alpha_{1,n}$, using McMahon's formula \cite{doi:10.1098/rspa.1918.0006}, the hardwall Regge trajectory behaves as $M_n^2\propto n^2$. Therefore, regarding the Regge trajectories for mesons, the geometric effect of intersecting a stack of D3-branes with another one of D7-branes living in a 10-dimensional Minkowski space separated by a distance $L$ is \emph{equivalent} to slicing the AdS$_5$ space with a hard cutoff. In both scenarios, the meson mass depends entirely on distance: in the D3/D7 case, $L$, and the hardwall model case, $z_\text{HW}$. And both scenarios have the same \emph{large $n$} behavior of the spectrum: $M_n^2\propto n^2$.

Thus, we can conclude that the bottom-up potential \eqref{D3_D7-pot} \emph{mimics} the spectroscopy of the top-down D3/D7 system.

\subsubsection{Fixing parameters}
For simplicity, we can write the D3/D7 trajectory for S-wave mesons ($l=0$) as 

\begin{equation}
    M_n^2=a(n+1)(n+2),\,\,\text{with:}\,a=\frac{4\,L^2}{R^4}.
\end{equation}

We focus our attention on light vector mesons. To do so, we identify the $\rho(770)$ meson as the ground state with $n=0$, and its mass to fix the energy scale $a$ as 

\begin{equation}\label{a_mass}
 M_{\rho(770)}^2 =2\,a,\,\,\,\text{thus:}\,\,\,a=0.30051\pm0.00018\,\text{GeV}^2,
\end{equation}

with $M_{\rho(770)}=0.77526\pm0.00023$ GeV \cite{Workman:2022ynf}.

This energy scale $a$ will allow us to compute the mass spectrum, thermal structure, and configurational entropy to characterize this \emph{reconstructed} bottom-up model. A plot of the reconstructed potential $V(z)$ using the fixed parameters is given in Fig. \ref{fig:one}. 
 
\subsubsection{Dilaton Reconstruction} 
\begin{figure}
\includegraphics[width=3.4 in]{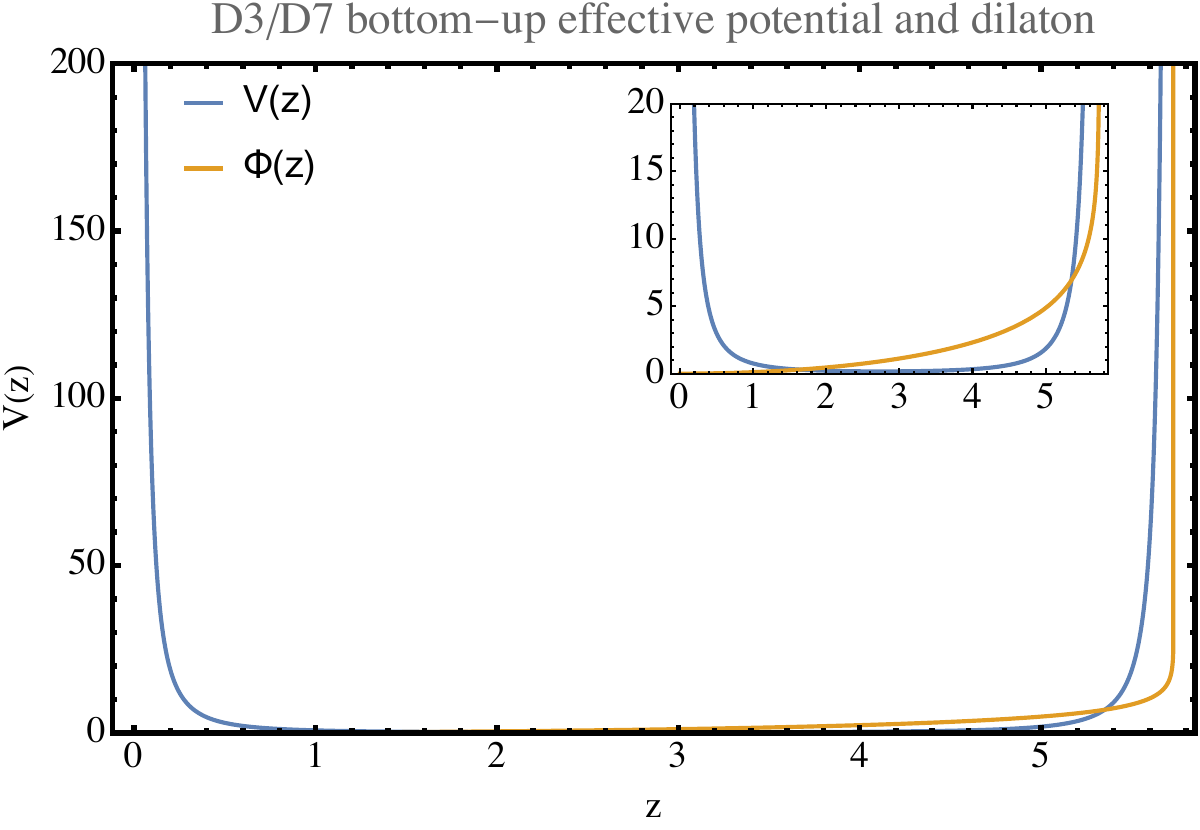}
\caption{Reconstructed bottom-up potential from the D3/D7 meson spectrum and the corresponding static dilaton. Notice that the dilaton diverges asymptotically at \emph{wall} located at $z=\pi/\sqrt{a}$. We have used the $\rho(770)$ meson mass to fix parameters according to the equation \eqref{a_mass}.  Experimental masses come from PDG \cite{Workman:2022ynf}.}
\label{fig:one}
\end{figure}

To find the dilaton field associated with the WKB-reconstructed potential, we will require as boundary conditions that $\Phi(z\to0)=0$ and 

\begin{equation}
    \Phi(z^*\to\infty)=2\,\int^{z^*}{dz\,\sqrt{V_\text{WKB}(z)}}, 
\end{equation}

\noindent with $z^*$ a turning point  (i.e., where $V(z^*) = M_n^2$ for a given state) in the asymptotic region. This set of conditions will allow us to solve the differential equation for the dilaton given in \eqref{dilaton-eng}. A plot of this dilaton field $\Phi(z)$ is depicted in Fig. \ref{fig:one}. 

Let us comment on the functional behavior of the dilaton. The dilaton in the region $z\to 0$ vanishes. However, in the vicinity of $z=\pi/\sqrt{a}$, the dilaton diverges. Thus, the dilaton is effectively \emph{confined spatially}. In the bulk action \eqref{action-mes}, this characteristic will cause the bulk dynamics to be restricted, i.e., the quantity $\pi/\sqrt{a}$ behaves as a wall. Thus, $a$ is a \emph{confinement scale} in the same sense as $\kappa$ in the softwall model \cite{Karch:2006pv}.

\subsection{Holographic spectrum of the Reconstructed Potential}
Let us focus on this part of the holographic spectrum associated with the potential \eqref{D3_D7-pot} when we consider $S$-wave vector mesons. 

As usual in the bottom-up approach, our starting point is to define the geometrical background. We will consider this system to be living in the Poincaré patch, given by the line element \eqref{AdS-metric}. Vector mesons in $S$-wave are dual to vector massless bulk fields, obeying the following Lagrangian density \eqref{lagrangian}:

\begin{equation}
    \mathcal{L}_\text{Mesons}=-\frac{1}{2\,g_5^2}\,\nabla_M\,\phi^{N}\,\nabla^M\,\phi_{N}
\end{equation}

\begin{figure}
\includegraphics[width=3.4 in]{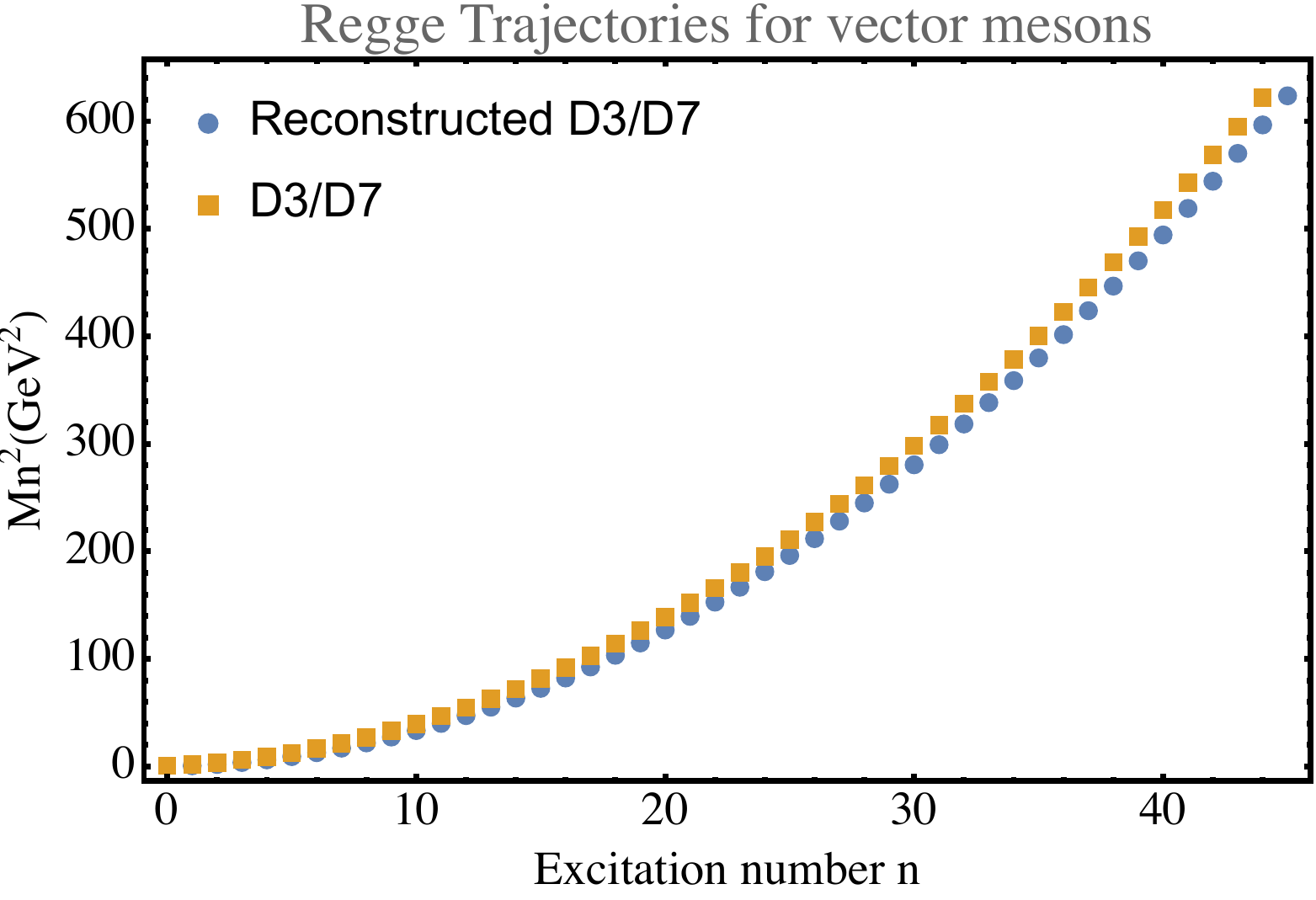}\\
  \includegraphics[width=3.4 in]{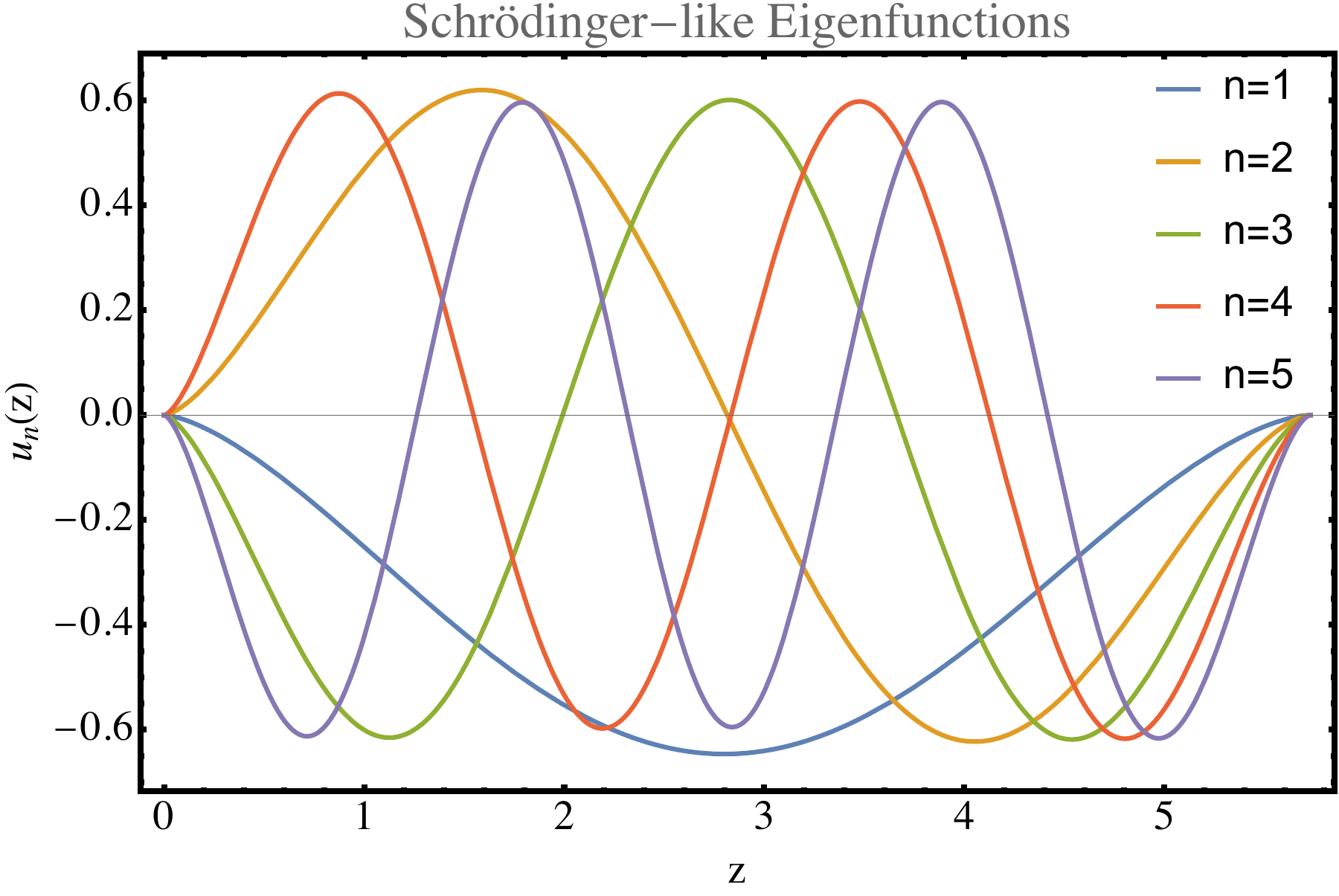}
\caption{The upper panel shows the radial Regge trajectories $M_n^2$ for vector mesons in the D3/D7 system \cite{Erdmenger:2007cm} and for the reconstructed one. The lower panel presents a plot of the ground state and the first four excited states for the reconstructed model.    }
\label{fig:four}
\end{figure}

Recall that the bulk mass is fixed to zero for the $S$-wave vector mesons: $M_5^2\, R^2=0$, since $\Delta=3$ in this case. Using this Lagrangian and the reconstructed dilaton plotted in Fig. \ref{fig:one}, we can compute the Sturm-Liouville equation \eqref{eq-sch-like}, given by   

\begin{equation}
    \partial_z\left[e^{-B(z)}\,\psi(z,q)\right]+M_n^2\,e^{-B(z)}\,\psi(z,q)=0,
\end{equation}

\noindent and the corresponding Schrödinger-like equation given by

\begin{equation}
    -u''(z)+V_{\text{D3/D7}}(z)\,u(z)=M_n^2\,u(z).
\end{equation}

\noindent where we have performed the standard Bogoliubov transformation $\psi(z)=e^{\frac{1}{2}B(z)}\,u(z)$. The plot for Schrödinger-like modes $u_n(z)$ and Regge trajectories for D3/D7 for the reconstructed WKB model is shown in Fig. \ref{fig:four}.

As we mentioned in the last section, for the D3/D7 system, we used the $\rho$ meson mass to fix the slope as $a=0.3$ GeV$^2$. This value was also used to fix the effective wall $z_\text{cutoff}$, according to the expression \eqref{cutoff}. The summary of the masses for the $\rho$ radial trajectory is given in Table \ref{tab:one}. We have compared the WKB-reconstructed outcomes with those for the hardwall and D3/D7 cases. In the reconstructed case, the ground state $\rho(770)$ emerges with a relative error smaller than $3\%$. Recall that the RKR formula does not give rise to an isospectral potential whose spectrum is $M_n^2$. The matching condition is performed at the large-$z$ limit \cite{MartinContreras:2023eft}. 

The hardwall model, which also exhibits quadratic behavior in $n$, increases faster than the D3/D7 and reconstructed models, since the behavior of the Bessel zeroes spectrum $\alpha_{1,n}$ determines the masses. For higher excitations, the quadratic behavior deviates from the expected linear tendency observed for light-unflavored mesons. 

\begin{table*}[t]
    \centering
    \begin{tabular}{c|c|c|c|c}
    \hline
    \multicolumn{5}{c}{\textbf{Light Vector mesons $I^G(J^{PC})=1^+(1^{--})$}}\\
    \multicolumn{5}{c}{\textbf{Masses in MeV}}\\
    \hline
    \textbf{State} & \textbf{Exp.}&\textbf{HWM} & \textbf{D3/D7} & \textbf{Reconstructed WKB}\\
    \hline \hline
    \textbf{$\rho(770)$}  &$775.26\pm0.23$ & $775.23\pm0.23^*$& $775.23\pm0.23^*$ & $796.02\pm0.01\, (2.7\%)$ \\
    \textbf{$\rho(1450)$}  &$1465\pm25$ & $1779.55\pm0.53\, (21\%)$& $1342.79\pm0.39\, (8\%)$ & $1367.69\pm0.02\, (7\%)$ \\
    \textbf{$\rho(1700)$}  &$1720\pm20$ & $2789.76\pm0.83 \,(62\%)$& $1898.99\pm0.56\, (10\%)$ & $1925.85\pm0.02\, (11\%)$ \\
    \textbf{$\rho(1900)$}  &$1880\pm10$ & $3801.32\pm1.13\,(102\%)$& $2451.59\pm0.73\,(30\%)$ & $2479.59\pm0.02\,(32\%)$ \\
    
    \end{tabular}
    \caption{Summary of vector meson masses for the $\rho$ trajectory calculated using the hardwall (HW), D3/D7, and reconstructed models. The asterisk (*) indicates the experimental mass used to fix parameters in each model. For the reconstruction, we use the same Regge slope value, $a=0.30051\pm0.00018$ GeV$^2$, used in the D3/D7 case. For the hardwall, we used $z_\text{HW}=3.10196\pm0.00092$ GeV, as given in eqn.  \eqref{HW-mass}.  Quantities inside parentheses correspond to relative errors with experimental data\cite{Workman:2022ynf}.}
    \label{tab:one}
\end{table*}

\section{Thermal Hawking-Page transition}\label{thermal}
A useful test is to analyze the thermal phases associated with this D3-D7-inspired solution \cite{Herzog:2006ra, Cai:2007zw}. In the AdS/CFT seminal works, Witten demonstrated that the thermodynamics of a gauge theory at the boundary is connected with the thermodynamics of AdS space. In particular, the Hawking-Page transition in the bulk geometry carries information about the confinement/deconfinement process \cite{Herzog:2006ra, Cai:2007zw}. The standard prescription is to compute the free energy in both solutions, the thermal AdS and the black hole one, AdS-BH. 

Thus, to study thermodynamics, we consider the Euclidean versions of the metrics. These solutions are given by

\begin{eqnarray}
    dS^2_\text{Th}&=&\frac{R^2}{z^2}
\left[d\tau^2+dz^2+d\vec{x}\cdot d\vec{x}\right]\\
dS^2_\text{BH}&=&\frac{R^2}{z^2}
\left[f(z)\,d\tau^2+\frac{dz^2}{f(z)}+d\vec{x}\cdot d\vec{x}\right]
\end{eqnarray}

\noindent with $f(z)=1-\frac{z^4}{z_h^4}$, and $z_h$ is the locus of the event horizon. The free energy is computed from the bulk action

\begin{multline}\label{HP-Acction}
  I=-\frac{1}{2\,\mathcal{G}^2}\int{d^5x\,\sqrt{g}\,e^{-\Phi(z)}\,(\,\mathcal{R}-\Lambda)}\\
  -\frac{1}{\mathcal{G}^{2}}\int_{\partial\mathcal{M}}d^{4}x \sqrt{h}  e^{-\Phi(z)} \mathcal{K},  
\end{multline}

\noindent evaluated on-shell using the thermal and AdS-BH solutions. We compute the free-energy difference as $\Delta\,F=F_\text{Th}-F_\text{BH}$ \cite{Braga:2022yfe}. 

\begin{figure}
  \includegraphics[width=3.4 in]{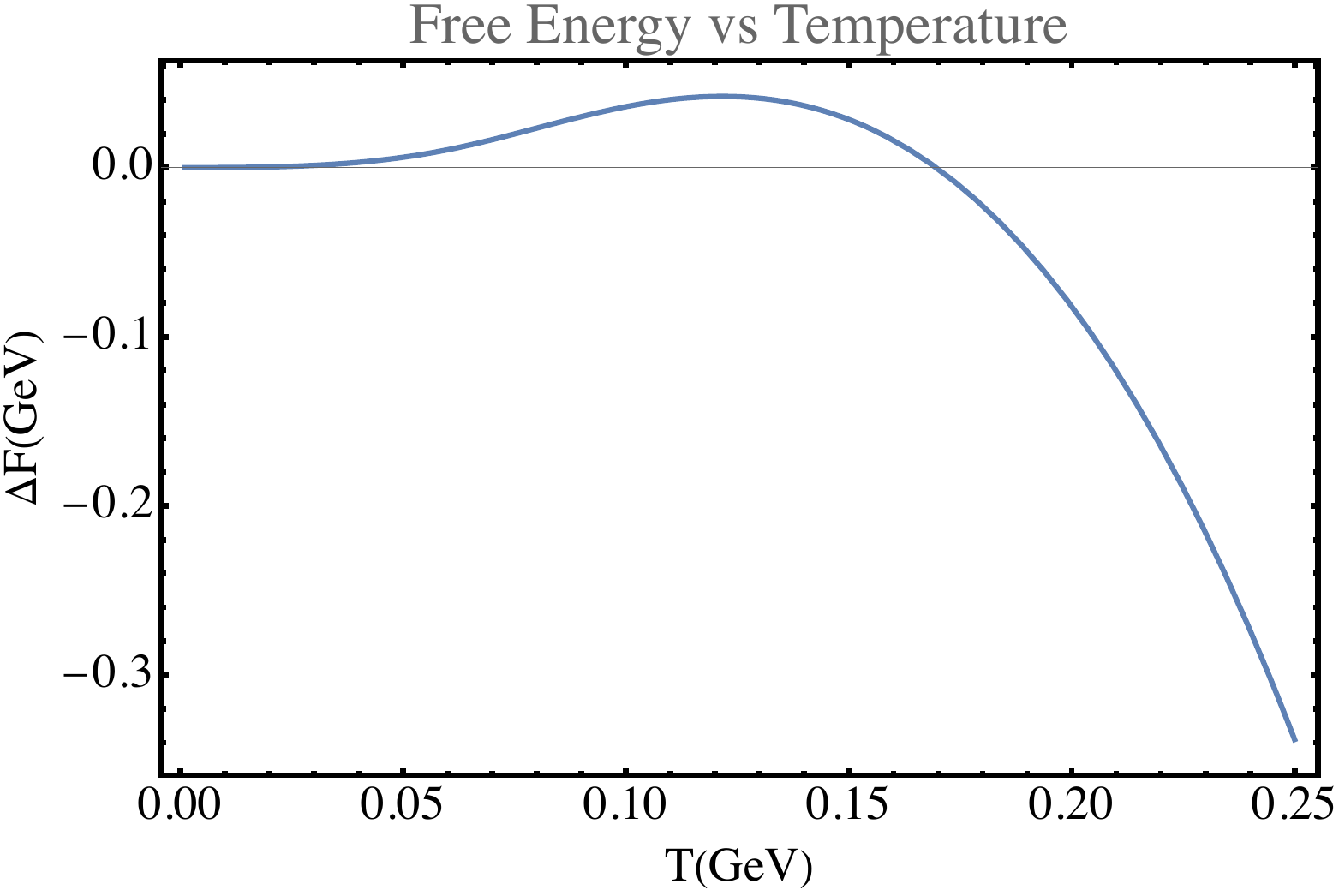}
\caption{Thermal Hawking-Page transition for the WKB-reconstructed dilaton $\Phi(z)$, with $a=0.3$ GeV$^2$. In this scenario, the confinement/deconfinement phase transition has a temperature of $169.6$ MeV.}
\label{fig:two}
\end{figure}

Notice that $\mathcal{G}^2$ contains the information of the five-dimensional Newton constant, $\mathcal{R}$ is the Ricci scalar, $\Lambda$ is the cosmological constant,  $\mathcal{K}$ is the extrinsic curvature scalar, and $h$ is the determinant of the boundary-induced metric. The last term in the action is necessary for regularization purposes \cite{BallonBayona:2007vp}. Figure \ref{fig:two} depicts a plot of the bulk free energy difference for the WKB-reconstructed dilaton. 

From the Hawking-Page analysis, and keeping in mind the relation between the wall locus and the lightest vector meson in the Regge trajectory exposed by Herzog, for this WKB model, we found that for the critical temperature, 

\begin{equation}
T_c=0.169\,\,\text{GeV}=0.218\,m_\rho,
\end{equation}

\noindent which is higher than the critical temperature of the hardwall model result, $0.1574\,m_\rho$. However, this critical temperature is lower than the softwall result $0.2459\,m_\rho$. See \cite{Herzog:2006ra} for further details. Thus, we can conclude that: 

\begin{equation}
    T_{c,\,\text{HWM}}<T_{c,\,\text{WKB}}<T_{c,\,\text{SWM}} 
\end{equation}

\noindent for the thermal confinement/deconfinement phase transition. 

In order to deliver a broader perspective on the deconfinement temperature we found above, we will compare with experimental and theoretical results summarized in Table \ref{tab:three}. Experimentally,  the deconfinement temperature is determined by analyzing particle production in heavy-ion collisions. Usually, the chemical freeze-out temperature (where inelastic collisions cease) is taken as the point at which hadronization begins. For example, the ALICE collaboration \cite{Baral:2026ohe} identifies the chemical freeze-out temperature around $150-160$ MeV.

In the case of lattice QCD, recent results tend to suggest that there is an identification between chiral restoration and deconfinement, which are expected to be close at zero baryonic chemical potential \cite{HotQCD:2018pds}. Using the \emph{center vortex geometry} \cite{Mickley:2024vkm}, two different scales emerge: $155$ MeV for chiral restoring and $321$ MeV for deconfinement. In the case of pure gluon systems, the transition is first-order, with a critical temperature of around $265$ MeV. Nevertheless, an important remark should be made: if the analysis is performed on Polyakov loops on a lattice with physical quarks, the deconfinement crossover has a pseudo-critical temperature in the $150-160$ MeV range \cite{Bazavov:2023xzm}.

When we consider other holographic bottom-up approaches, the deconfinement temperature is identified with the Hawking-Page phase transition in the dual gravity. Thus, estimates vary depending on the model and input parameters. Thus, since the criterion for fine-tuning parameters is spectroscopical, it depends on the hadronic species analyzed \cite{Afonin:2014jha}.

\begin{table*}[t]
    \centering
    \begin{tabular}{c|c|c}
         \hline
         \multicolumn{3}{c}{\textbf{Deconfinement Temperature}}\\
         \hline
         \textbf{Source} & $T_c$(MeV) & \textbf{Method}\\
         \hline
         \hline
        ALICE \cite{Baral:2026ohe,Flor:2021olm} & $155-165$ & Thermal-FIST framework\\
        HotQCD\cite{HotQCD:2018pds}& $156.5\pm1.5$  & HISQ  \\
        Gavai et al. \cite{Gavai:2024mcj} &$158.7^{+2.6}_{-2.3}$ &(2+1) flavor QCD with M\"obius domain-wall\\
        Mickley et al. \cite{Mickley:2024vkm} & $321\pm6$& Center vortex analysis   \\
        Herzog \cite{Herzog:2006ra} & $191$ & Softwall Model\\
        Herzog \cite{Herzog:2006ra} & $125$ & Hardwall Model\\
        Kim et al. \cite{Kim:2007em} & $100-130$ & Hardwall with quark backreaction\\
        Afonin et al. \cite{Afonin:2014jha} & $260$ & Softwall with planar gluodynamics\\
        Afonin \cite{Afonin:2018era} & $235-305$ & Isospectral Softwall for scalar glueball\\
        Chen \cite{Chen:2024mmd} & $147$ & EMD-Data driven $2+1$ model\\
        \textbf{This work} & $169$ & Reconstructed WKB. 
    \end{tabular}
    \caption{Summary of deconfinement temperatures for different analyses (Experimental, non-holographic, and holographic). }
    \label{tab:three}
\end{table*}

\section{Configurational Entropy and Stability}\label{DCE}

Configurational entropy (CE) is associated with the various arrangements (or microstates) that a specific macrostate can assume. Consequently, a higher CE indicates a greater number of potential microstate configurations. In thermodynamic terms, this entropy is associated with the work done by a system while remaining in a fixed spatial configuration, independent of any energy exchange.      

In information theory, configurational entropy (CE) is an important metric for assessing the relationship between the informational content of physical solutions and their corresponding equations of motion (e.o.m.). CE is a logarithmic measure of the spatial complexity inherent in localized solutions given a specific energy content. Consequently, it quantifies the informational content embedded in the solutions to a particular set of equations of motion. 

In more precise terms, CE can be viewed as an indicator of the degree of information required to describe how spatially localized the solutions to e.o.m. in the bulk are. Generally, dynamic solutions result from extremizing an action. Thus, CE measures the information available within these solutions.

In the framework of AdS/CFT, the holographic perspective on configurational entropy has been explored within both bottom-up and top-down AdS/QCD models \cite{Bernardini:2016hvx}. The concept was initially introduced for the hadronic states in \cite{Bernardini:2016qit, Braga:2017fsb, Colangelo:2018mrt, Ferreira:2019nkz, Ferreira:2020iry, daRocha:2021ntm} and the associated references. Regarding the stability of heavy quarkonium, DCE was employed as a methodological tool to investigate thermal characteristics within a colored medium \cite{Braga:2018fyc}, considering the influence of magnetic fields \cite{Braga:2020hhs} and finite density \cite{Braga:2020myi}. The study in \cite{Braga:2020opg} used CE to examine the holographic deconfinement phase transition within bottom-up AdS/QCD frameworks. Furthermore, recent research, as highlighted in \cite{MartinContreras:2022lxl}, employs configurational entropy to analyze holographic stability in light nuclides. CE was also used to describe stable non-$Q\bar{Q}$ hadronic structures \cite{MartinContreras:2023oqs}, isospectrality \cite{MartinContreras:2023eft}, and $\Sigma$ baryons using bottom-up holography \cite{Guo:2024nrf}.

For systems exhibiting instability, we can posit that configurational entropy is a tool for analyzing their degree of stability. Generally, the hadronic mass can be expressed as a function that increases with configurational entropy, $S_{\mathrm{CE}}$. Furthermore, based on Heisenberg's uncertainty principle, it is possible to relate the decay width $\Gamma$ to the hadron mass, e.g., $\Gamma \sim M_{n} \sim S_{\mathrm{CE}}^{\gamma}$, with $\gamma>0$, as referenced in  \cite{MartinContreras:2023eft}. Thus, for a holographic model that intends to describe hadrons, \emph{a good phenomenological test} is to observe increasing configurational entropy with the excitation number. 

For vector hadrons, the recipe for configurational entropy $s_\text{CE}$ can be summarized as follows: 

\begin{figure}
  \includegraphics[width=3.4 in]{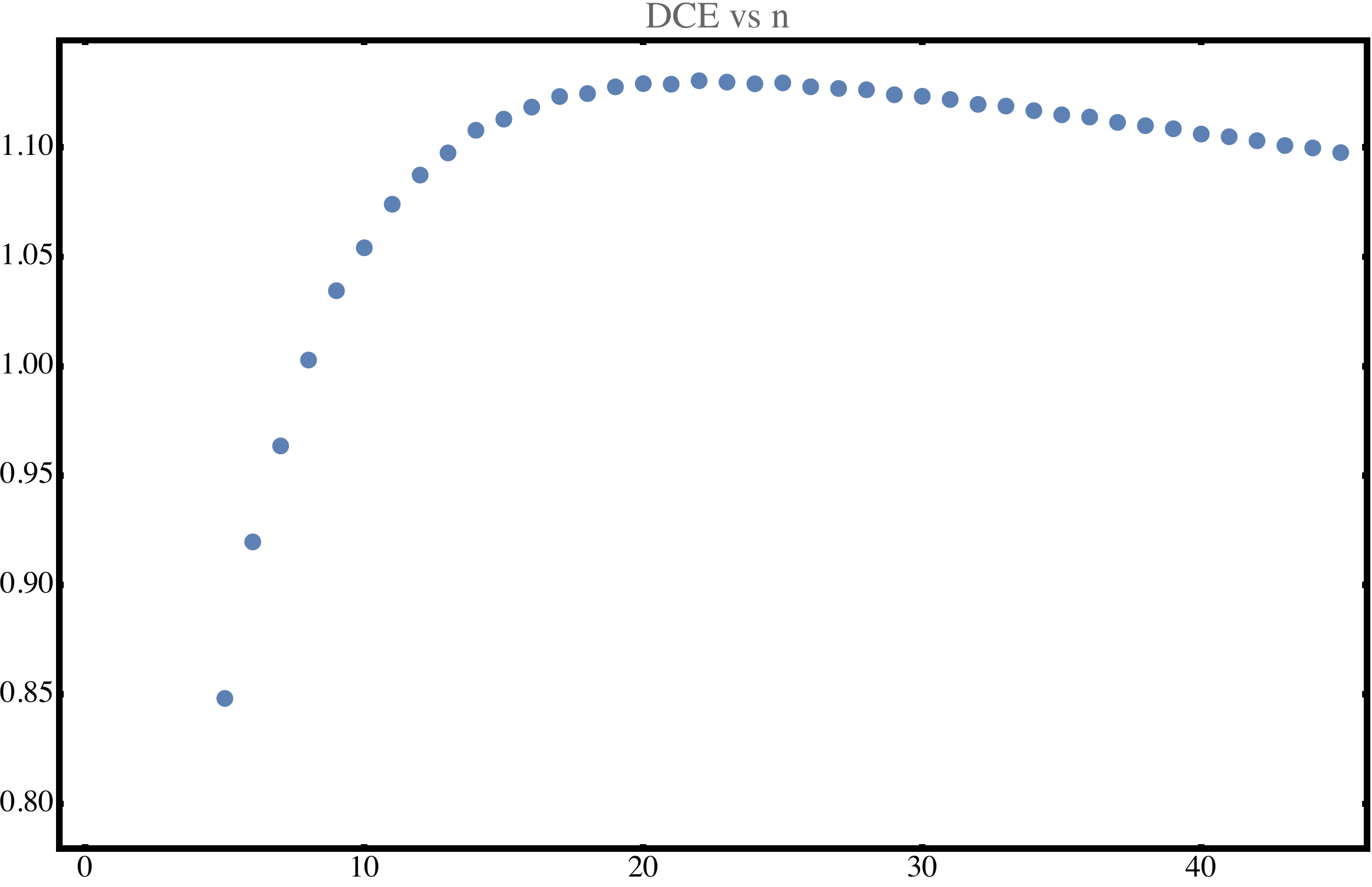}
\caption{Differential configurational entropy as a function of the excitation number $n$ for vector mesons in the context of WKB-reconstructed potential. For high enough $n$, DCE decreases.}
\label{fig:three}
\end{figure}

\begin{enumerate}
    \item Start from the bulk action \eqref{action-mes} computing the equations of motion associated with the bulk fields dual to hadrons. 
    \item compute the on-shell energy-momentum tensor $T_{MN}$:

    \begin{equation}\label{Energy-momentum}
T_{MN}=\frac{2}{\sqrt{-g}}\,\frac{\partial\left[\sqrt{-g}\,\mathcal{L}_\text{Hadron}\right]}{\partial\,g^{MN}}.    
\end{equation}

\item Extract the $T_{00}$ component, defined as the bulk energy density $\rho(z)$. For bulk vector fields, $\rho(z)$ has the form:
\begin{multline}\label{energy-density}
 \rho(z)=\frac{e^{-B\left(z\right)}}{2}\left(\frac{z}{R}\right)^3\times\\
\left\{ \left[\frac{1}{\mathcal{K}^2}\left(M_n^2\,\psi_n^2+\psi_n'^2\right)-\frac{M_5^2\,R^2}{z^2}\psi_n^2\right] \right\}\,\Omega,
\end{multline}
\noindent where $\Omega$ is a constant factor resulting from plane wave and polarization contractions. This factor cancels in the subsequent normalization steps for the modal fraction and is therefore omitted hereafter.

\item Fourier-transform the energy density $\rho(z)$:
\begin{equation}
    \bar{\rho}(k) = \int_0^{\infty}d\,z\, e^{ik\,z}\rho(z)
\end{equation}
\noindent and compute the \emph{modal fraction} as
\begin{equation}
   f(k) =\frac{ |\bar{\rho}(k)|^2}{\int dk |\bar{\rho}(k)|^2}.
\end{equation}
Recall that $\rho(z)\in L^2(\mathbb{R})$, and also has information on how energy is localized in the bulk. Thus, it indirectly measures how normalizable modes are well localized in the AdS space. Thus, the modal fraction measures the spread of the energy in the bulk. 
\item Finally, compute the \emph{differential configurational entropy} with the following prescription:
\begin{equation}
S_{DCE}=-\int dk \,\tilde{f}(k) \log \,\tilde{f}(k)
\end{equation}

\noindent where $\tilde{f}\left(k\right)=f\left(k\right)/f\left(k\right)_\text{Max}$.  Notice that we normalize the modal fraction with $f\left(k\right)_\text{Max}$.
\end{enumerate}

We computed the differential configurational entropy (DCE) in natural units for the WKB-reconstructed system across the first 45 states. The calculation is depicted in Fig. \ref{fig:three}. Notice that, as we expect from phenomenology, DCE increases for the first sixteen states. When we increase the excitation number, DCE decreases, as expected for the hardwall model \cite{MartinContreras:2023eft}. This suppression effect is associated with the presence of the wall, imposed by the $\tan^{2}(\sqrt{a}z/2)$ term in the holographic potential, which causes the \emph{DCE overlapping} of the highest states with the lower ones. This behavior is a known characteristic of confining geometries with a sharp cutoff \cite{Braga:2017fsb}, reflecting the finite number of degrees of freedom accessible below the wall. It serves as a diagnostic tool, distinguishing this class of models from those with truly infinite, linear Regge trajectories.

\section{Conclusions}\label{concl}
This work is motivated by the goal of obtaining a confining holographic potential from a given eigenvalue spectrum. A noteworthy application of this investigation is to examine the mass spectrum of hadrons derived from top-down models and subsequently determine the corresponding bottom-up confining potential, thereby elucidating the relationship between these two frameworks for the description of hadrons. Notably, in the context of bottom-up AdS/QCD, the mass spectrum is determined by the asymptotic behavior of the holographic potential at large $z$. Consequently, applying the WKB method to the Regge trajectory can derive the corresponding potential and dilaton field. 

While the quantum-mechanical inverse problem does not guarantee a unique solution, the potential we derive successfully reproduces the key spectroscopic features of the D3/D7 system and mirrors the hardwall geometry. 

Recall that in the D3/D7 case, the meson masses arise from the consistent truncation of the geometry generated by the intersecting D3 and D7 stacks. Meson states correspond to fluctuations in the D7-brane shape (gauge fields) on the D7-Brane Worldvolume. These fluctuations propagate along the intersection and are localized in the extra dimensions by an effective \emph{potential well}. Thus, the dual meson states live in the closed internal geometry formed by the D7-brane embedding.

 On the other hand, in the hardwall model, the idea is to cut AdS space with a topological defect (such as a brane) that imposes Dirichlet or Neumann boundary conditions. This process allows the emergence of bound bulk states dual to hadrons. Normalizable solutions live in a closed compact AdS slice. Thus, we conclude that, in the hardwall model, hadrons are also a consequence of geometric deformations.

Therefore, at least at an intuitive level, the D3/D7 system and the hardwall model have similar behavior regarding the Mass spectrum and its origin. To establish a complete equivalence, we require that each possible hadronic observable that can be calculated in both models should behave similarly. 

The methodology developed here, however, extends beyond this specific top-down example. The primary strength of the RKR prescription is its applicability to experimental Regge trajectories, enabling a more data-driven derivation of confining holographic potentials.

We also analyzed the phase transition between confinement and deconfinement within the reconstructed geometry. Our findings indicate that the critical temperature is approximately $169$ MeV, higher than that of the conventional hardwall model and smaller than that of the softwall model with a quadratic static dilaton. 

Finally, we performed a configurational entropy analysis to assess the feasibility of the reconstructed model for describing hadrons. The analysis reveals that the configurational entropy does not increase with $n$, contrary to initial expectations, as it exhibits a local maximum. Nonetheless, for the lowest modes ($n<16$), the configurational entropy increases with $n$, reflecting a hardwall-like behavior. Consequently, this reconstructed WKB model can potentially descriptively represent light meson phenomena in a holographic manner.

\begin{acknowledgments}
M. Fujita would like to thank A. Karch for the helpful discussion. M. A. Martin Contreras would like to acknowledge the financial support provided by the National Natural Science Foundation of China (NSFC) under grant No. 12350410371. A. Vega is partially supported by the Centro de Física Teórica de Valparaíso (CeFiTeV).
\end{acknowledgments}

\appendix

\section{Sumary of Rydberg-Klein-Rees procedure}\label{app-1}
For completeness, we summarize the standard RKR procedure and its adaptation to the holographic case in the appendices below.

The Rydberg \cite{Rydberg1, Rydberg2}-Klein \cite{Klein}-Rees \cite{Rees} method is a semiclassical inversion procedure for the Schrödinger equation that allows one to obtain a bound-state potential from an energy spectrum.

Let us consider the radial Schrödinger equation as the starting point:

\begin{equation}
- \frac{\hbar^{2}}{2 m} \frac{d^{2} u_{E}}{d\,r^2} + V(r) u_{E} = E u_{E}.
\end{equation}

According to the WKB approach, the quantization condition is given by:

\begin{equation}
n + \frac{1}{2} = \frac{1}{\pi} \sqrt{\frac{2 m}{\hbar^{2}}} \int^{r_{2}}_{r_{1}} \sqrt{E - V(r)} dr,
\end{equation}

\noindent where $n$ is the radial quantum number and $r_{1,2}$ refers to the turning points.

The main result in the RKR method encompasses two integrals involving the turning points (for details of the inversion procedure, see the appendix in \cite{RKRref}):

\begin{equation}\label{RKR1}
r_{2} - r_{1} = 2 \sqrt{\frac{\hbar^{2}}{2 m}} \int^{n}_{n_{min}} \frac{d n'}{\sqrt{E(n) - E(n')}} 
\end{equation}
\begin{equation}\label{RKR2}
\frac{1}{r_{2}} - \frac{1}{r_{1}} = 2 \sqrt{\frac{\hbar^{2}}{2 m}} \int^{n}_{n_{min}} \frac{B_{n'}\,d n'}{\sqrt{E(n) - E(n')}}.
\end{equation}

In molecular physics, the first equation is called the \emph{vibrational RKR equation}. The second one is the \emph{rotational RKR equation}, with $B_{n}$ given by:

\begin{equation}
B_{n} = \frac{\partial E(n,l)}{\partial [l(l+1)]} \biggr|_{l=0}
\end{equation}

\noindent with $l$ defined as the angular momentum number. Therefore, from a well-known spectrum, the RKR equations give us a collection of turning points, which can be interpolated to obtain a potential in this spectrum.

\section{Rydberg-Klein-Rees procedure for holography}\label{app-2}

In AdS/QCD models, we can always write the equation for AdS modes that describe hadrons as:

\begin{equation}
    -u''(z)+V(z)\,u(z)=M_n^2\,u(z),
\end{equation}

\noindent with the holographic potential defined in terms of the dilaton and the AdS warp factor as: 

\begin{multline}\label{holographic-pot-1}
   V(z)=\frac{M_5^2\,R^2}{z^2}-\frac{\beta\,(2-\beta)}{4\,z^2}-\frac{\beta}{2\,z}\Phi'(z)\\
   +\frac{1}{4}\Phi'(z)^2-\frac{1}{2}\Phi''(z).
\end{multline}

Here, the WKB quantization condition reads as:

\begin{equation}
n + \frac{1}{2} = \frac{1}{\pi} \int^{z_{2}}_{z_{1}} \sqrt{M^{2} - V(z)} dz,
\end{equation}
where $n$ is the radial quantum number and $z_{1,2}$ are the turning points.

We use $V^*$ to denote the potential calculated using WKB from the full bottom-up holographic potential $V(z)$, which can be written as:

\begin{equation}
    V(z)=\frac{\beta(\beta-2)+4\,M_{5}^2\,R^2}{4\,z^2}+V^*(z).
\end{equation}

Compared to the expression \eqref{holographic-pot-1}, the potential $V^*(z)$ is solely dependent on the dilaton field, and for large $z$. It also provides the primary contribution for higher states. Thus, in this case, we can approximate \eqref{holographic-pot-1} by $V^*(z)$ with turning points in $z=0$ and $z=z(V^*)$ to calculate the mass spectrum for radial excitations. This fact significantly simplifies our procedure since we can employ the RKR method to obtain $V^*(z)$ by merely considering the equation 
\eqref{RKR1}, which is applicable in our case as:

\begin{equation}\label{RKR1holo}
z(V^*) = 2  \int^{n(M^{2})}_{n_{min}} \frac{d n'}{\sqrt{M^{2}(n) - M^{2}(n')}}.
\end{equation}

Since $M^2=M^2 (n)$, we can change the integration variable in the last integral by considering the turning points $M^2(n) = V^*$ and $M^{2}_{n_{min}} \sim 0$. Focusing on the large $z$ (higher radial excitations), we obtain the following integral:

\begin{equation}\label{RKR1holo2}
z(V^*) = 2  \int^{V^*}_{0} \frac{d M^{2}}{\frac{d M^{2}}{dn} \sqrt{V^{*} - M^{2}}}.
\end{equation}

Therefore, using a single RKR equation allows one to deduce the asymptotic behavior of the potential near the turning points, thereby enabling the extraction of a dilaton.

\bibliography{apssample}
\end{document}